\documentclass[preprint,showpacs,preprintnumbers,amsmath,amssymb]{revtex4}

\usepackage{graphicx}
\usepackage{dcolumn}
\usepackage{bm}
\usepackage{epsfig}

\def\e{\begin{equation}}
\def\f{\end{equation}}
\def\=#1{\overline{\overline #1}}

\def\-#1{{\bf #1}}
\def\.{\cdot}

\begin{document}

\title{Experimental Demonstration of a Structured Material with Extreme Effective Parameters at Microwaves}

\author{M\'{a}rio G. Silveirinha$^1$}
\altaffiliation[]{Electronic address: mario.silveirinha@co.it.pt}
\author{Carlos A. Fernandes$^2$}
\author{Jorge R. Costa$^{2,3}$}
\author{Carla R. Medeiros$^{2}$}
\affiliation{$^1$University of Coimbra, Department of Electrical
  Engineering-Instituto de Telecomunica\c{c}\~{o}es, 3030 Coimbra,
  Portugal \\
$^2$Technical University of Lisbon, Instituto Superior
T\'{e}cnico-Instituto de Telecomunica\c{c}\~{o}es, 1049-001
Lisbon,Portugal\\
$^3$Instituto Superior de Ci\^{e}ncias do Trabalho e da Empresa,
Departamento de Ci\^{e}ncias e Tecnologias da Informa\c{c}\~{a}o,
1649-026 Lisboa, Portugal}

\date{\today}

\begin{abstract}
Following our recent theoretical studies [M. G. Silveirinha, C. A.
Fernandes, Phys. Rev. B, 78, 033108, 2008], it is experimentally
verified that an array of crossed metallic wires may behave as a
nonresonant material with extremely large index of refraction at
microwaves, and may enable the realization of ultra-subwavelength
waveguides.
\end{abstract}

\pacs{42.70.Qs, 78.20.Ci, 41.20.Jb, 78.66.Sq} \maketitle

The design of novel materials with unusual electromagnetic
properties has received a lot of interest in recent years, mainly
because of the great potentials of structures with simultaneously
negative permittivity and permeability \cite{Pendry_PerfectLens}.
However, materials with extreme parameters or extreme properties,
such as near-zero-permittivity,  extreme anisotropy, or
very-large-permittivity may also have interesting applications in
many problems such as tunneling through narrow channels and bends
\cite{SilvEnghetaPRL}, subwavelength imaging \cite{SWIWM},
realization of magnetic materials in the visible domain
\cite{metameta}, and cloaking \cite{PendryCloak}.

\begin{figure}[thb] \centering
\epsfig{file=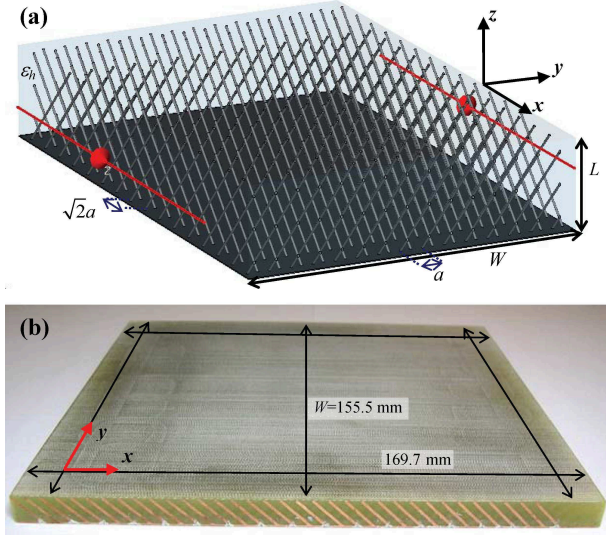, width=8cm} \caption{(Color online) (a) A
grounded metamaterial slab is placed in between two horizontal
dipole antennas. The structured material is formed by an array of
crossed metallic wires parallel to the \emph{xoz} plane, and are
tilted by $\pm$45$^o$  with respect to the interfaces. The wires are
embedded in a host material with relative permittivity
$\varepsilon_h$. (b) Photo of the fabricated prototype formed by 97
FR4 boards printed with metallic strips.} \label{geomwires}
\end{figure}

In a recent work \cite{MarioEVL}, we have reported that a composite
material formed by crossed metallic wires may have an anomalously
strong interaction with electromagnetic waves, and may enable the
realization of materials with extremely large positive index of
refraction. The geometry of a grounded slab of the structured
material is depicted in the panel (a) of Fig. \ref{geomwires}. It
was theoretically predicted in \cite{MarioEVL} that such grounded
slab may interact with a wave with electric field polarized along
the $x$-direction as a material with very large $positive$
permittivity. This result is rather surprising since systems formed
by long metallic wires are typically thought as materials with
$negative$ permittivity. Here, we experimentally verify  the results
of Ref. \cite{MarioEVL}, and demonstrate that the proposed material
may enable the realization of ultra-subwavelength waveguides.

To this end, a prototype of the structured material was fabricated
using a layer by layer design and printed circuit techniques [panel
(b) of Fig. \ref{geomwires}]. In our planar design the metallic
wires are replaced by printed metallic strips. The material is
formed by an array of 97 layers of 1.6mm thick FR4 boards, for which
the experimentally determined permittivity around the 1.3 GHz design
frequency was $\varepsilon_h = 4.3$ with loss tangent $\tan
\delta=0.02$. The boards of FR4 have been printed with parallel
metallic strips with width $w_s = 0.83$mm. The metallic strips are
tilted by $\pm 45^{o}$ with respect to the interfaces, and thus the
strips in adjacent FR4 boards are mutually orthogonal. The lattice
constant is $a=3.2$mm, and the height of the substrate is
$L=2.6a=8.32$mm. As illustrated in panel (a) of Fig.
\ref{geomwires}, the metallic strips are supposed to be connected
with good ohmic contact to a metallic ground plane. In practice,
such configuration may be difficult to reproduce using planar
technology. To avoid this difficulty, we have mimicked the ground
plane by printing an additional metallic strip running along the
$x$-direction  at the lower edge of each FR4 board. This ensures
good ohmic contact between all the strips in the same board. In
addition, to ensure ohmic contact between the stacked FR4 boards,
each FR4 board was perforated at 17 specific spots next to the lower
edge [these spots can be identified in the lower panel of  Fig.
\ref{geomwires}], and copper wires (running along the $y$-direction)
were inserted through these spots and soldered to the boards.
Finally, the bottom face of the fabricated prototype was coated with
a silver paint. Since the spacing between the copper wires is very
small as compared to the wavelength, such configuration effectively
behaves as a continuous ground plane.
\begin{figure}[tb] \centering
\epsfig{file=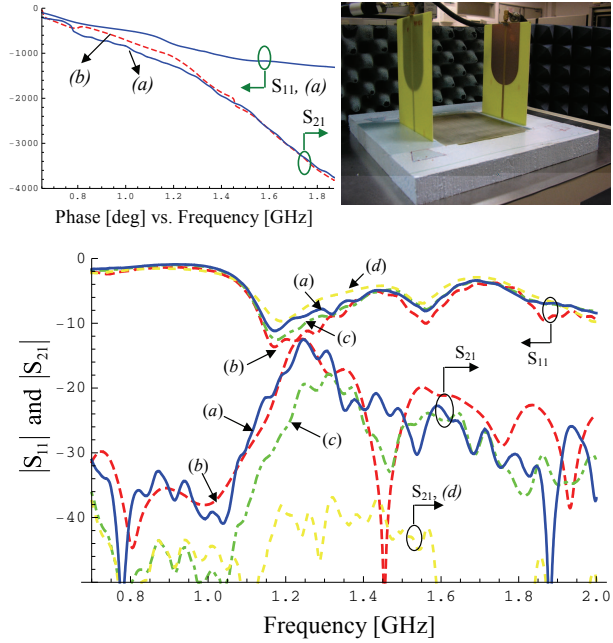, width=8cm} \caption{(Color online) Lower
panel: Amplitude of the S-parameters in dB: (a) measured results.
(b) CST Microwave Studio full wave simulations. (c) measured results
when an absorber is placed in between the two antennas. (d) same as
(c) but the prototype is rotated by $90^o$. Top left panel: Phase
for cases (a) and (b). Top right panel: photo of the experimental
setup.} \label{spars}
\end{figure}

In order to characterize the properties of the material, we have
investigated the propagation of transverse electric (TE) guided
modes in the structured substrate. It is well known that a grounded
dielectric slab with relative permittivity $\varepsilon$ only
supports guided modes when $L>{{\lambda _0 } \mathord{\left/
 {\vphantom {{\lambda _0 } {4\sqrt {\varepsilon  - 1} }}} \right.
 \kern-\nulldelimiterspace} {4\sqrt {\varepsilon  - 1} }}$. Thus, a
grounded substrate with the same permittivity as the FR4 boards only
supports TE-guided modes for frequencies larger than 4.98GHz. As
theoretically shown in \cite{MarioEVL}, the presence of the crossed
wire mesh may result in a dramatic increase of the effective
permittivity, and consequently in a significant reduction of the
TE-modes cut-off frequency. In order to experimentally verify this
property and measure the dispersion characteristic of the guided
modes, the structured substrate was excited by an horizontal dipole
antenna oriented along the $x$-direction and placed 5mm away from
the metamaterial substrate. The transmitted field was measured by an
identical dipole placed at the other end of the substrate [top panel
of Fig. \ref{geomwires}]. Actually, the antennas used in the
experiment were not wire dipoles, but instead were printed in FR4
boards as well. Each printed dipole is connected to the vector
network analyzer (VNA) through a ``balun'' that makes the proper
transition between the VNA coaxial cable and the co-planar printed
line that feeds the dipole. In order, that the ``baluns'' are not
detuned by the presence of the structured substrate it is convenient
that the dielectric boards with the feeding circuit are oriented in
such a way that they are normal to the $y$-direction [top right
panel of Fig. \ref{spars}]. The free-space resonance frequency of
the printed dipoles is approximately 1.2GHz.

The phase and amplitude of the measured S-parameters are depicted in
the top left and lower panels of Fig. \ref{spars}, respectively
[curves (a)]. The calibration planes are at the antenna SMA
connectors. Notice that $\left| {S_{11} } \right|$ is a measure of
the fraction of power accepted by the input antenna, whereas $\left|
{S_{21} } \right|$ is a measure of the coupling between the two
dipoles. It is important to mention that even if there is an
excellent coupling between the two dipoles the level of $\left|
{S_{21} } \right|$ does not need to be close to 0dB, since part of
the power accepted by the input antenna is either radiated as space
wave, or dissipated in the dielectrics and metals, and only the
remaining power is received by the test antenna, being the main
propagation mechanism a guided wave. It is seen in Fig. \ref{spars}
that for frequencies below 1.1GHz the level of $\left| {S_{21} }
\right|$ is below -30dB, showing that for these frequencies the
coupling between the dipoles is very low. This is understandable
since for low frequencies the TE-guided modes are cut-off. However,
around 1.25GHz there is dramatic increase of the level of $\left|
{S_{21} } \right|$ which approaches -12dB, suggesting that the
structure may support a guided mode above 1.25GHz, notwithstanding
the electrically small thickness of the substrate
($L=\lambda_0/29$). This property stems from the anomalously high
effective permittivity of the structured material \cite{MarioEVL}.
For frequencies above 1.35GHz, the $\left| {S_{21} } \right|$ level
remains significant, even though it deteriorates with increasing
frequency, in part because the dipoles were tuned to 1.2GHz, and
thus they tend to radiate poorly for frequencies relatively far from
1.2GHz. We have experimentally verified with a different set of
printed dipoles tuned to 1.5GHz, that the onset of propagation of
guided modes is still at 1.25GHz, whereas the $\left| {S_{21} }
\right|$ level may be improved around 1.5GHz [not shown here].
Another reason for the deterioration of the $\left| {S_{21} }
\right|$ level is that slightly above the cut-off frequency the
guided modes become very confined inside the structured material,
and thus it is difficult to launch the guided mode from the air
region. This is supported by the extremely fast variation of the
$S_{21}$ phase slightly above 1.25GHz, suggesting that the guided
mode has an extremely short wavelength, and consequently is highly
confined to the structured material because of the very large
effective permittivity.

The reported experimental data concur very well with results
obtained with a fullwave commercial electromagnetic simulator
\cite{CST2006} [curves (b) in Fig. \ref{spars}]. In the numerical
simulations the ground plane was assumed continuous. The excellent
agreement between the numerical and experimental results demonstrate
that our implementation of the ground plane effectively mimics an
ideal continuous ground plane.

\begin{figure}[thb] \centering
\epsfig{file=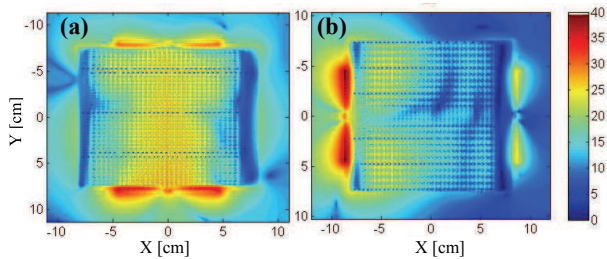, width=8cm} \caption{(Color online) CST
Microwave Studio simulations at 1.25GHz for the total electric field
amplitude at the mid-height of the artificial substrate for dipole
antennas (a) parallel to the $x$-axis, (b) parallel to the
$y$-axis.} \label{Efield}
\end{figure}

In order to show that the main propagation mechanism between the two
dipoles is indeed a guided wave, we have placed an 80mm thick
microwave absorber block in between the two antennas, 7mm above the
prototype. The purpose of the absorber was to block the space wave
radiated by the dipole. The corresponding measured S-parameters are
shown in Fig. \ref{spars} [curves (c)]. It is seen that except near
1.25GHz, the $\left| {S_{21} } \right|$ level is little affected by
the presence of the absorber, consistently with our expectations
that the dominant propagation mechanism is the guided mode. The drop
of the $\left| {S_{21} } \right|$ close to the cut-off is
understandable, since at the cut-off frequency most of the energy of
the guided mode propagates in the air region, and thus is also
blocked by the absorber. To demonstrate in a conclusive manner, that
the enhancement of the $\left| {S_{21} } \right|$ level is due to
the emergence of guided modes, we have rotated the prototype by
$90^o$ so that the dipole antennas become orthogonal to the FR4
boards with the printed strips. In this case, it is expected that
the interaction of the metallic strips with the dipoles is very
weak, and thus the prototype should behave in the same manner as an
homogeneous grounded substrate with the same permittivity as the FR4
boards. The measured results corresponding to this scenario are
depicted in the lower panel of Fig. \ref{spars} [curves (d)], and
demonstrate that indeed the $\left| {S_{21} } \right|$ level is
close to -40dB, consistently with the fact that a homogeneous
dielectric substrate with $\varepsilon_h = 4.3$, only supports
TE-guided modes above 4.98GHz. The results obtained with CST
Microwave Studio completely confirm these findings. In Fig.
\ref{Efield} we plot the total electric field amplitude at 1.25GHz
along the mid-height cut of the structured substrate for the cases
in which the dipoles are parallel to either the $x$-axis or to the
$y$-axis. In these simulations the region in between the printed
dipoles was not blocked by the microwave absorber. The plots
demonstrate how in the case in which the dipoles are parallel to the
FR4 boards (panel (a)), the wave is able to penetrate into the
structured substrate permitting a good coupling between the two
dipoles. This situation contrasts markedly with the case in which
the dipoles are perpendicular to the FR4 boards (panel (b)), for
which the guided mode cannot be launched and the wave cannot
propagate along the substrate.
\begin{figure}[thb] \centering
\epsfig{file=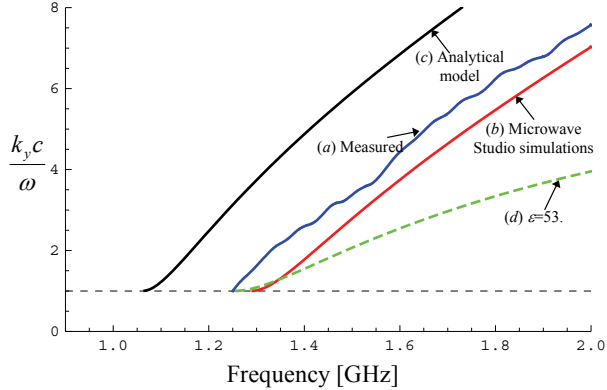, width=8cm} \caption{(Color online)
Dispersion characteristic of the guided mode supported by the
structured material: (a) measured result. (b) full wave simulation
with \cite{CST2006}. (c) theoretical result. (d) dispersion for a
dielectric slab with $\varepsilon=53$. } \label{dispSWs}
\end{figure}

It is possible to calculate the dispersion characteristic of the
guided modes from the experimental data. A straightforward analysis
shows that the propagation constant of the guided mode, $k_y$, is
such that, $k_y \approx \frac{1}{W}\left[ {2\arg \left( {1 - \rho }
\right) - \arg \left( \tau  \right) + C} \right]$, where $W$ is the
width of the substrate [see Fig. \ref{geomwires}], $C$ is a constant
chosen so that $k_y  = \omega/c$ at the cut-off frequency of the
guided mode, and $\rho$ and $\tau$ are the $S_{11}$ and $S_{21}$
parameters referred to the input ports of the printed dipoles,
respectively [these parameters are obtained by proper deembedding of
the experimental data obtained at the antenna SMA connectors]. The
extracted dispersion characteristic of the guided mode is depicted
in Fig. \ref{dispSWs} (blue curve). It was assumed in the extraction
procedure that the cut-off frequency is 1.25GHz, consistent with the
results of Fig. \ref{spars}. In Fig. \ref{dispSWs} we have also
plotted the dispersion characteristic calculated with CST Microwave
Studio \cite{CST2006}, obtained by analyzing a single cell of the
periodic substrate (red curve), as well as the dispersion
characteristic obtained using the analytical model described in our
previous work \cite{MarioEVL} (black curve) \footnote{The model
proposed in \cite{MarioEVL} assumes that the inclusions are wires
and not strips; we have taken the radius of the
wires equal to $R=\pi w_s$, 
so that the perimeter of the cross-section of the inclusions is the
same.}. It is seen that the results obtained with the different
methods concur very well, and that the value of $k_y c / \omega$ may
be significantly larger than unity for frequencies slightly above
the cut-off frequency. This supports the fact the guided mode
becomes highly confined in the structured material. In curve (d) of
Fig. \ref{dispSWs}, we show the dispersion characteristic of a
grounded dielectric slab, with the same cut-off frequency as the
prototype. The permittivity of the equivalent dielectric slab is
$\varepsilon = 53.0$, which demonstrates how the crossed wire mesh
may in fact enhance in a dramatic manner the electric properties of
the host medium. As discussed in \cite{MarioEVL}, much larger values
of the permittivity may be easily obtained by increasing the density
of wires. It is interesting to note that the slope of $k_y c /
\omega$ vs. frequency is much larger for the structured material
than for the equivalent dielectric slab. This property is a
consequence of the fact that the response of the bulk structured
material is anisotropic, even for propagation in the $yoz$ plane.
Quite remarkably, this property enables a greater confinement of the
electromagnetic fields inside the metamaterial slab, and thus
suggests interesting applications of the proposed structure as an
ultra-subwavelength waveguide. This work is supported in part by
Funda\c c\~ao para a Ci\^encia e a Tecnologia, grant number No.
POSC/EEACPS/61887/2004.


\end{document}